\documentclass[12pt]{article}

\usepackage{amsmath,epsfig,ulem,cite,tabularx}

\newcommand{\msn}[1]{m_{\tilde{\nu}_{#1}}}
\newcommand{\mseR}{m_{\tilde{e}_{\rm R}}}

\newcommand{\mst}{m_{\tilde{t}_1}}

\newcommand{\cha}{\tilde{\chi}}
\newcommand{\neu}{\tilde{\chi}^0}
\newcommand{\mcha}[1]{m_{\tilde{\chi}^\pm_{#1}}}
\newcommand{\mneu}[1]{m_{\tilde{\chi}^0_{#1}}}

\def\mathswitch#1{\relax\ifmmode#1\else$#1$\fi}
\def\mathswitchr#1{\relax\ifmmode{\mathrm{#1}}\else$\mathrm{#1}$\fi}
\newcommand{\PW}{\mathswitchr W}
\newcommand{\PZ}{\mathswitchr Z}

\newcommand{\Pt}{\mathswitchr t}

\newcommand{\MW}{\mathswitch {m_\PW}}
\newcommand{\MZ}{\mathswitch {m_\PZ}}

\newcommand{\mt}{\mathswitch {m_\Pt}}

\newcommand{\tev}{\,\, \mathrm{TeV}}
\newcommand{\gev}{\,\, \mathrm{GeV}}

\newcommand{\SLASH}[2]{\makebox[#2ex][l]{$#1$}/}

\newcommand{\Eslash}{\SLASH{E}{.5}\,}

\newcommand{\anc}{\rule{0mm}{0mm}}
\newcommand{\lesim}{\,\raisebox{-.1ex}{$_{\textstyle <}\atop^{\textstyle\sim}$}\,}
\newcommand{\gesim}{\,\raisebox{-.3ex}{$_{\textstyle >}\atop^{\textstyle\sim}$}\,}
\newcommand{\knickpfeil}{\;\raisebox{1.12ex}{$\lfloor$} \!\!\! \to}

\newcommand{\mycaption}[1]{\caption{\sl #1}}

\begin{document}
\thispagestyle{empty}

\def\thefootnote{\fnsymbol{footnote}}

\begin{flushright}
FERMILAB--PUB--06--289--T
\\
ZH--TH 18/06
\end{flushright}

\vspace{1cm}

\begin{center}

{\Large\bf Collider searches and cosmology\\[1ex] in the MSSM with heavy
scalars}
       \\[3.5em]
{\large  
M.~Carena$^{1}$,
A.~Freitas$^{2}$}

\vspace*{1cm} 

{\sl
$^1$ Fermi National Accelerator Laboratory, Batavia, IL 60510-0500, USA

\vspace*{0.4cm}

$^2$ Institut f\"ur Theoretische Physik,
        Universit\"at Z\"urich, \\ Winterthurerstrasse 190, CH-8057
        Z\"urich, Switzerland
}

\end{center}

\vspace*{2.5cm}

\begin{abstract}

In a variety of supersymmetric extensions of the Standard Model, the scalar
partners of the quarks and leptons are predicted to be very heavy and beyond
the reach of next-generation colliders. For instance, the realization of
electroweak baryogenesis in supersymmetry requires new sources of CP-violation,
which can only be naturally accommodated with electric dipole moment constraints
if the first and second generation scalar fermions are beyond the TeV scale.
Also in focus-point supersymmetry and split supersymmetry the scalar fermions
are very heavy. In this work, the phenomenology of scenarios with electroweak
baryogenesis and in the focus point region at the LHC and ILC is studied,
which becomes challenging due to the presence of heavy scalar fermions.
Implications for
the analysis of baryogenesis and dark matter are deduced. It is found that
precision measurements of superpartner properties allow an accurate
determination of the dark matter relic density in both scenarios, while
important but only incomplete information about the baryogenesis mechanism can
be obtained.

\end{abstract}

\def\thefootnote{\arabic{footnote}}
\setcounter{page}{0}
\setcounter{footnote}{0}

\newpage


\section{Introduction}

Supersymmetry is one of the most compelling extensions of the Standard Model,
with the possibility to explain the stabilization of the electroweak symmetry
breaking scale, the existence of dark matter in the universe and 
the generation of the baryon-anti\-baryon-asym\-metry at the electroweak phase
transition (electroweak baryogenesis).
The existence of dark matter in the universe has been firmly established by
various experiments, and has been measured precisely by the Wilkinson Microwave
Anisotropy Probe (WMAP) \cite{wmap}, in agreement with the Sloan Digital Sky
Survey (SDSS) \cite{sdss}, $\Omega_{\rm CDM} h^2 =
0.1106^{+0.0056}_{-0.0075}$ at the 68\% C.L. Here $\Omega_{\rm CDM}$ is the
ratio of the dark matter energy density to the critical density $\rho_{\rm c} =
3 H_0^2 / (8\pi G_{\rm N})$, where $H_0 = h \times 100$ km/s/Mpc is the Hubble
constant and $G_{\rm N}$ is Newton's constant. In the Minimal Supersymmetric
extension of the Standard Model (MSSM), the stable lightest neutralino is an
attractive candidate for dark matter.

The process of electroweak baryogenesis requires a sufficiently strongly first
order electroweak phase transition, $v(T_{\rm c})/T_{\rm c} \gesim 1$, where
$v(T_{\rm c})$ denotes the Higgs vacuum expectation value at the critical
temperature $T_{\rm c}$ \cite{phasetran}. Moreover, new sources of CP violation
in addition to the CKM matrix phase are necessary \cite{CPSM}. Loop effects of
light scalar top quarks (stops) can induce a strongly first order electroweak
phase transition, thus generating the out-of-equilibrium condition for
electroweak baryogenesis \cite{CQW,EWBG,Carena:1997ki,EWBG2}.  In addition, CP
violation in the chargino sector of the MSSM can explain the magnitude of the
baryon asymmetry.

The parameter space for successful electroweak baryogenesis in the MSSM is
already highly constrained. The lightest stop mass needs to be in the range of
roughly 115 to 140 GeV, whereas the Higgs boson involved in the electroweak
phase transition must be lighter than
about 120 GeV \cite{CQW,EWBG,Carena:1997ki,EWBG2,stop}. Furthermore, a
CP-violating
phase in the chargino sector is highly constrained by bounds on electric dipole
moments \cite{edm1,edm2}.

Most MSSM scenarios predict a dark matter density larger than the measured
value. However, the requirement of a light stop from the baryogenesis mechanism
suggests that co-annihilation between the stop and the lightest neutralino can
bring down the relic density to the proper region. It has been shown that  the
co-annihilation is efficient for mass differences between the light stop and
the lightest neutralino, that are smaller than about 30 GeV
\cite{Balazs:2004bu}.

Light stops can be searched for at the Tevatron, with a reach up to stop
masses of about 170 GeV for 2--4 fb$^{-1}$ of integrated luminosity
\cite{Demina:1999ty}. However, if the stop-neutralino mass difference is less
than 30--50 GeV, the stop signal cannot be identified at the Tevatron, due to
the reliance on a trigger for missing transverse energy. At the next generation
of colliders, the situation looks more promising. The Large Hadron Collider
(LHC), depending on the region of parameter space, can see a signal from stops
in gluino cascades, while a future international $e^+e^-$ linear collider (ILC)
has excellent capabilities to discover and analyze light stops
\cite{stopsLHC, stopsLC,stopsnew,stop}.

In order to allow sufficiently large CP violating phases for baryogenesis in
the chargino sector, but evade current experimental electric dipole moment
bounds \cite{edmexp}, without invoking miraculous cancellations, the sfermions
of the first two generations need to be heavier than a few TeV, thus
effectively decoupling from collider experiments. This situation has grave
consequences for the investigation of supersymmetry at the LHC and ILC. While
the lightest Higgs boson, and possibly some of the heavier Higgs boson, are
within experimental reach as in other MSSM scenarios, only few supersymmetric
particles are likely to be kinematically accessible in this scenario:
neutralinos and charginos, the light stop and potentially the gluino and some
sbottom and stau
states. Thus it becomes much more difficult to identify and measure the
properties of these particles. A similar situation occurs in focus-point
\cite{focusp} and split supersymmetry scenarios \cite{splitsusy}, where all
sfermions are predicted to be very heavy.

In this work, it is investigated how a scenario with heavy sleptons and squarks
can be studied at the LHC and the ILC. The analysis is carried out in detail
for an electroweak baryogenesis scenario and for a focus-point scenario. Split
supersymmetry is not studied separately, but the conclusions are very similar
to the focus-point case. After reviewing the definition of the relevant
parameters as well as experimental and theoretical constraints in
section~\ref{sec:not}, the phenomenology of the baryogenesis scenario is
discussed in section~\ref{sec:bgen}, while section~\ref{sec:fp} is devoted to
the focus-point scenario. The results are based on a phenomenological analysis,
including backgrounds and systematic experimental limitations, but without a
realistic detector simulation and based on tree-level formulae only. For both
scenarios, the cosmological implications to be gained from collider results are
derived for elucidating the nature of baryogenesis and dark matter. For the
main part of the work, it is assumed that both electron and positron
beam polarization are available at the ILC. In section~\ref{sec:pol} it
is studied how the results change without positron polarization. Finally,
the conclusions are given in section~\ref{sec:concl}.


\section{Baryogenesis and focus-point scenarios in the MSSM}
\label{sec:not}

\paragraph{Baryogenesis} in the MSSM requires an additional source of CP
violation beyond the Standard Model CKM matrix. Within the MSSM, the dominant
source are chargino and neutralino loops, with a contribution proportional to
${\rm Im}\{\mu M_{1,2}\}$ \cite{Carena:1997gx,Carena:2002ss}. 
Here $\mu$ is the Higgs/higgsino parameter and $M_2$ and $M_1$ are the soft
SU(2) and U(1) gaugino parameters, respectively. To generate a
sufficiently large baryon asymmetry, the charginos are required to be
relatively light, $\mcha{1} \sim {\cal O}(\mbox{a few 100 GeV})$. In addition,
the CP-violating phase needs to be sizable, $\arg(\mu M_2) \gesim 0.1$
\cite{Carena:2002ss}.

A very large CP-violating phase, on the other hand, is restricted by
experimental bounds on the electric dipole moments of the electron, neutron and
$^{199}$Hg nucleus \cite{edmexp}. The leading contributions from one-loop
sfermion-gaugino
loops \cite{edm1,edm2} become small for large masses of the first two
generation sfermions of several TeV or for large cancellations between the
sfermion mass and mixing parameters. Here, to avoid the constraints from the
one-loop contributions, the sleptons and squarks of the first two generations
are assumed to have masses of about 10 TeV. Then the leading contributions from
supersymmetric particles to the electric dipole moments come from two-loop
diagrams. For moderate values of $\tan\beta < 10$, the ratio of the vacuum
expectation values of the two Higgs doublets, and the CP-violating phase,
it has been shown \cite{morr,caltech} that the two-loop effects are in
agreement with current experimental constraints.

Generically, at the LHC many supersymmetric
particles could be observed in the decay chain of squarks, whereas at ILC
selectron production would be one of the cleanest and most precise testing
grounds for supersymmetry. However, in the baryogenesis
motivated scenario, the large mass of the first and second generation
sfermions effectively
decouples them from observable processes. This has a large impact on the
prospects for future collider experiments, since the squark and slepton
production channels mentioned above are not available. Likewise, the dynamics of
baryogenesis and
dark matter annihilation are governed by the light particles in the MSSM
spectrum, which include, beside the stops, the charginos and neutralinos.
Therefore the following analysis focuses on the phenomenology of the chargino
and neutralino particles.

The spectrum of the two charginos and
four neutralinos is described by the parameters $\mu$, $M_1$, $M_2$ and
$\tan\beta$.
A CP-violating phase in $M_2$ can always be transferred into the $\mu$
parameter by means of a unitary transformation. In principle, there can also be
non-trivial phases in the gaugino parameters $M_1$ and $M_3$. While the effect
of a phase of $M_3$ on electroweak baryogenesis is small, a complex phase of
$M_1$ could have interesting consequences, but is not investigated further here.
Therefore in the following all gaugino
soft parameters are assumed real, while the generation of the baryon asymmetry
is connected with a phase in the $\mu$ parameter,
$
\mu = |\mu| \times e^{i \phi_\mu}.
$

For definiteness, the specific MSSM parameter point BGEN will be considered in the
following, as defined in the appendix. Since the allowed parameter space for
baryogenesis in the MSSM is already highly constrained by experimental bounds,
this particular scenario is typical for the general MSSM baryogenesis case.

In the MSSM Higgs sector, the tree-level masses of the CP-even neutral Higgs
bosons $h^0$ and $H^0$ and the charged scalar $H^\pm$ can be expressed through
the gauge boson masses, the mass of the pseudo-scalar Higgs boson, $m_{\rm
A^0}$, and $\tan\beta$. The Born relations are however significantly modified
by radiative corrections, with dominant effects originating from top and stop
loops. The mass of the Higgs state related to electroweak symmetry breaking (in
most cases $h^0$) is very sensitive to the stop spectrum. In order to
be consistent with the bound $m_{\rm h^0} \gesim 114.4 \gev$ from direct
searches
at LEP \cite{lephbound} and with one light stop state, the heavier stop mass
has to be above about 1 TeV and the trilinear coupling $A_t$ has to be
sizable \cite{Carena:1997ki}.  Constraints from electroweak precision
data, in particular the $\rho$ parameter,
are satisfied when the light stop is mainly right-chiral. This is
naturally achieved for values of the stop supersymmetry breaking parameters
$m_{\rm\tilde{Q}_3}^2 \gesim 1 \tev^2$ and $m_{\rm\tilde{U}_3}^2 \lesim 0$,
respectively. The stop mixing parameter $X_t = \mu \cot \beta - A_t$ is bounded
from below by the Higgs boson mass constraint from LEP and from above by the
requirement of the strength of the first order electroweak phase transition,
leaving the allowed range $0.3 \lesim |X_t|/m_{\rm\tilde{Q}_3} \lesim 0.5$
\cite{Carena:1997ki}. The value of $\tan\beta$ is also constrained to the range
$5 \lesim \tan\beta \lesim 10$, with the upper bound stemming from present
electric dipole moment limits and the lower bound is again related to the LEP
Higgs mass limit. The latter can be weakened for large values of
$m_{\rm\tilde{Q}_3}$ of several TeV.
Also values of $m_{\rm A^0}$ larger than about 200 GeV are preferred in order to
be compatible with the electric dipole moment bounds.

The MSSM Higgs masses with CP violation have been calculated  including
complete one-loop and leading two-loop corrections, see {\it e.g.} 
Ref.~\cite{higgsmass}. In this work, however, the process of baryogenesis at
the electroweak phase transition is computed with the program of
Refs.~\cite{Carena:2002ss,morr}, which includes only one-loop corrections to
the zero temperature Higgs potential. Since the allowed mass range for the
Higgs boson  is constrained by the mechanism of electroweak baryogenesis, for
consistency the Higgs mass is determined by the minimization of the one-loop
effective potential. This implies that only one-loop corrections are included
in the calculation of the Higgs mass as well. Although when including two-loop
corrections, the values of the fundamental parameters will change for given
values of $m_{\rm h^0}$ and $v(T_{\rm c})/T_{\rm c}$, the correlation between
$m_{\rm h^0}$ and $v(T_{\rm c})/T_{\rm c}$ is expected not to be strongly
modified\footnote{An analysis including two-loop corrections to the effective
potential is in progress \cite{tleff}.}.

\paragraph{Focus point} supersymmetry \cite{focusp} was suggested to solve the
supersymmetric flavor and CP-problems by raising the masses of all scalars to
several TeV. It was observed that for certain parameter combinations, the Higgs
parameters have an infrared quasi-fixed point, thus making the weak scale soft
parameter $m_{H_{\rm u}}$ of one of the Higgs doublets highly insensitive to
the value of the other scalar masses. In this way, naturalness is preserved
even for very large sfermion masses.

Here, as a concrete example, the focus point scenario LCC2, as defined in the
appendix, will be analyzed in detail. The collider phenomenology of this
scenario has been studied previously in Ref.~\cite{FP}, and cosmological
implications were discussed in Ref.~\cite{peskin}. In this report, the
phenomenological discussion of the earlier works is extended, and the scenario
compared to the baryogenesis case.


\section{Baryogenesis scenario}
\label{sec:bgen}

The LHC experiments will be able to probe a light Higgs boson with
Standard-Model-like couplings to the gauge bosons, as required by electroweak
baryogenesis.
The potential for discovery of light stops has been
studied in detail for the LHC \cite{stopsLHC} and ILC
\cite{stopsLC,stopsnew,stop}. If gluinos are not too heavy, the discovery of
light stops in gluino decays via the process $pp \to \tilde{g} \tilde{g} \to t
t\, \tilde{t}^* \tilde{t}^*$, $\bar{t} \bar{t}\, \tilde{t} \tilde{t}$ has been
shown to be effective for small stop-neutralino mass differences
\cite{stopsLHC}. Stops can be discovered with more than five standard
deviations through the decay of gluinos if the gluino mass is below 900 GeV,
the stop-neutralino mass difference about 40 GeV, and all other squarks
relatively heavy. It was also shown \cite{stopsLHC} that if the other squarks
are not much heavier than 1 TeV, even in the stop-neutralino co-annihilation
region with $\mst - \mneu{1} \lesim 30$ GeV, stop discovery is possible in the
decay of those squarks. However, further studies are needed to establish whether
the stop-neutralino
co-annihilation region can be explored at the LHC for first and second
generation squarks with masses of several TeV. In this parameter region, the
decay products of the stops are very soft and thus escape direct detection, so
that only the final state products of the two top quarks from gluino decay are
observable. Nevertheless, the signature of same-sign top quarks and large
missing energy is likely sufficient for the discovery of a new physics signal,
even if the stops are not directly identified. When discovered, the analysis of
the kinematical decay distributions of the stops at the LHC can be used to
extract information about the relationship between stop, gluino and lightest
neutralino masses \cite{stopsLHC}, but an independent determination of the stop
or neutralino mass seems difficult.

At the ILC, light stops can be discovered for stop-neutralino mass differences
down to about 5 GeV, independent of other MSSM parameters \cite{stop}. In
addition, precise measurements of the stop mass and mixing angle can be
performed, with a typical stop mass error of about 1\%, assuming 80\% beam
polarization for electrons and 50\% for positrons.

\subsection{Chargino and neutralino studies at the LHC}
\label{sec:lhc}

In this kind of scenarios, the analysis of neutralinos and charginos is very
difficult. Supersymmetric particles are mainly produced through decay cascades
of gluino and squarks, which can have large production cross-sections at the
LHC. If however all squarks except the light stop are rather heavy, only the
stop will be produced at sizeable rates and it will dominate the decays of the
gluinos as well. However, a stop in accordance with electroweak baryogenesis is
so light that it typically can only decay into the lightest neutralino, so that
no information about the other neutralinos and charginos can be gained from
stop processes. 

The only sfermion whose mass is not constrained by theoretical or experimental
bounds in the MSSM baryogenesis scenario, is the right-chiral sbottom quark. 
The left-chiral sbottom is connected through SU(2) symmetry to the left-chiral
stop, which needs to be very massive to satisfy the LEP Higgs bound
\cite{lephbound}. The right-chiral sbottom, on the other hand, directly decays
into a bino neutralino, with negligible branching ratios into other neutralinos
or charginos. Note that the light sbottom state is expected to be almost
completely right-chiral in this scenario, since large mixing between left- and
right-chiral sbottom states is suppressed when the light mass eigenstate is
supposed to be much lighter than the heavy state.

Finally, charginos and neutralinos can also be
produced directly through Drell-Yan-type processes. However, direct production processes of
neutralino pairs have too small rates compared to
the backgrounds at the LHC \cite{tadas}. On the other hand, the production of a
light chargino with a neutralino can have sizeable rates, but in a baryogenesis
scenario, the chargino will mainly decay into the light stop. In the sample
scenario BGEN (see appendix), the chargino branching fraction into stops and
bottom quarks is more than 99.9\%. As a result, the production of mixed
neutralino-chargino pairs will lead to at most one lepton plus jets in the
final state, instead of the typical tri-lepton signature for mSUGRA scenarios.
Unfortunately, the one lepton plus jets signature is totally swamped by
$W$-boson background.

Thus the only observable supersymmetric particles at the LHC are, depending on
the region of parameter space, the gluino, the light stop, potentially the light
sbottom, and the missing energy signature of the lightest neutralino.


\subsection{Chargino and neutralino studies at the ILC}
\label{sec:ilc}

In the given scenario, stops, charginos and neutralinos can be studied
precisely at a future linear collider. The prospective measurements for the
stop are studied in detail in Ref.~\cite{stop}. Here the chargino and
neutralino phenomenology is investigated. The characteristic feature of the
baryogenesis scenario is that while the charginos and neutralinos can be light,
the first generation sleptons are heavy. A similar situation arises in focus
point supersymmetry and has been studied in previous works \cite{FP, FP2}. Here
the study of Ref.~\cite{FP2} is extended by including more observables,
simulations for signal and background processes and by considering a
CP-violating phase in the system.

In general, future collider experiments will not provide enough independent
measurements to extract all chargino and neutralino mass and mixing parameters
in a model-independent way. Therefore any attempt to fully reconstruct a
supersymmetric model from experimental data will rely on assuming some
structure for that model. In particular, in the MSSM, the chargino and
neutralino masses and couplings depend on five unknown parameters, 
namely $M_1$, $M_2$, $|\mu|$, $\phi_\mu$
and $\tan\beta$. In the MSSM, these parameters can be related to production
cross-sections and masses. Thus with a sufficient amount of measurements for
chargino and neutralino masses and cross-sections, all the underlying
parameters can be extracted.

For the sample scenario BGEN (see appendix), the chargino and neutralino masses at
tree-level amount to
\begin{equation}
\begin{aligned}
\mneu{1} &= 106.6 \gev, & \mneu{2} &= 170.8 \gev, & \mcha{1} &= 162.7 \gev, \\
\mneu{3} &= 231.2 \gev,	& \mneu{4} &= 297.7 \gev, & \mcha{2} &= 296.2 \gev. \\
\end{aligned}
\end{equation}
At the ILC with $\sqrt{s} = 500$ GeV, many neutralino and chargino states are
accessible, see Tab.~\ref{tab:ilcXsec}.
\renewcommand{\arraystretch}{1.2}
\begin{table}[tp]
\begin{center}
\begin{tabular}{|r||rrr|}
\hline
$e^+e^- \to \neu_i \neu_j$ & $\neu_i = \neu_2$ & $\neu_3$ & $\neu_4$ \\
\hline \hline
$\neu_j = \neu_1$ & 0.7 & 25.1 & 0.07 \\
$\neu_2$ & 0.4 & 62.0 & 0.005 \\
$\neu_3$ && --- & --- \\
$\neu_4$ &&& --- \\
\hline \hline
$e^+e^- \to \cha^\pm_i \cha^\mp_j$ & $\cha^\pm_i = \cha^\pm_1$ & $\cha^\pm_2$ & \\
\hline \hline
$\cha^\mp_j = \cha^\mp_1$ & 665 & 28 & \\
$\cha^\mp_2$ & & --- & \\
\hline
\end{tabular}
\end{center}
\vspace{-1em}
\mycaption{Tree-level production cross-sections in fb at $\sqrt{s} = 500$ GeV
with unpolarized beams
for the reference point BGEN.}
\label{tab:ilcXsec}
\end{table}
As evident from the table, the most promising processes are $e^+e^- \to \cha^+_1
\cha^-_1$, $e^+e^- \to \neu_1 \neu_3$ and $e^+e^- \to \neu_2 \neu_3$. The
production of mixed charginos pairs, $e^+e^- \to \cha^\pm_1
\cha^\mp_2$, also has a sizeable cross-section, but is overwhelmed by background
from $e^+e^- \to \cha^+_1\cha^-_1$.

The relevant decay processes are summarized in Tab.~\ref{tab:decays}.
\renewcommand{\arraystretch}{1.2}
\begin{table}[tb]
\begin{center}
\begin{tabular}{|c||c|c|r@{\:}lr|}
\hline
Sparticle & Mass $m$ [GeV] & Width $\Gamma$ [GeV]
          & \multicolumn{3}{c|}{Decay modes} \\
\hline \hline
$\neu_1$ & $106.6$ & --- & \multicolumn{2}{c}{---} & \\
$\neu_2$ & $170.8$  & $0.00002$ &
        $\neu_2$ & $\to Z^* \, \neu_1$ & 100\% \\
$\neu_3$ & $231.2$  & $0.11$ &
        $\neu_3$ & $\to Z \, \neu_1$ & 98\% \\
        &&&& $\to h^0 \, \neu_1$ & 2\% \\
$\neu_4$ & $297.7$  & $0.86$ &
        $\neu_4$ & $\to Z \, \neu_1$ & 1\% \\
        &&&& $\to Z \, \neu_2$ & $\ll 1$\% \\
        &&&& $\to Z \, \neu_3$ & 1\% \\
        &&&& $\to W^\pm \, \cha^\mp_1$ & 94\% \\
        &&&& $\to h^0 \, \neu_1$ & 4\% \\
        &&&& $\to h^0 \, \neu_2$ & $\ll 1$\% \\
\hline
$\cha_1^\pm$ & $162.7$  & $0.24$ &
        $\cha^+_1$ & $\to \tilde{t} \, \bar{b}$ & 100\% \\
$\cha_2^\pm$ & $296.2$  & $4.2$ &
        $\cha^+_2$ & $\to \tilde{t} \, \bar{b}$ & 84\% \\
        &&&& $\to W^+ \, \neu_1$ & $\ll 1$\% \\
        &&&& $\to W^+ \, \neu_2$ & 10\% \\
        &&&& $\to Z \, \cha^+_1$ & 6\% \\
        &&&& $\to h^0 \, \cha^+_1$ & $\ll 1$\% \\
\hline
\end{tabular}
\end{center}
\vspace{-1em}
\mycaption{Tree-level masses, widths and main branching ratios of the
neutralino and chargino states at Born level
for the reference point BGEN.}
\label{tab:decays}
\end{table}
While the two lightest neutralinos dominantly decay into (virtual) $Z$ bosons,
the lightest chargino will decay into the light stop and a bottom. For light
stops, the by far dominant decay mode is the loop induced process $\tilde{t} \to
c \, \neu_1$ \cite{Hikasa:1987db}.

\paragraph{Chargino \boldmath $\cha_1^+$}
\anc\\
Production of light charginos, $e^+e^- \to \cha_1^+ \cha_1^-$, has a large
cross-section of 665 fb at $\sqrt{s} = 500$ GeV. It can be further increased by
using beam polarization, $P(e^+)$/$P(e^-)$ =
$+50$\%/$-80$\%, where $+$ stands for right-handed, $-$ for
left-handed polarization, yielding $\sigma[e^+e^- \to \cha_1^+ \cha_1^-]
= 1760$ fb.

The dominant decay chain $e^+e^- \to \cha_1^+ \cha_1^- \to \tilde{t}
\tilde{t}^* \, b \bar{b} \to c \bar{c} \, b \bar{b} \, \neu_1 \, \neu_1$ leads
to two charm jets and two bottom jets plus missing energy in the final state.
The most important Standard Model backgrounds arise from production of vector
bosons in pairs (where the missing energy is generated by mismeasurements) and
triples (where neutrino decays can lead to missing energy), as well as
$t\bar{t}$ production. In addition, one needs to consider supersymmetric
background from neutralino production.

Signal and background events are simulated with the Monte-Carlo methods from
Ref.~\cite{slep}, including full tree-level matrix elements and Breit-Wigner
propagators for resonant intermediate particles. The processes are generated on
the parton level. Jet broadening through parton shower and detector effects are
parameterized by smearing functions with lepton and jet energy uncertainty taken
from \cite{tesladet}. Jets overlapping within a cone with $\Delta R =
\sqrt{(\phi_1-\phi_2)^2 + (\eta_1-\eta_2)^2} < 0.3$ are combined into one jet,
where $\phi_i$ and $\eta_i$ are the azimuthal angle and rapidity of jet $i$.
Similarly, a lepton lying within a jet is combined into the jet. Leptons and
jets outside the central region of the detector have a higher likelihood of
mistag and get inflicted by large two-photon background. Therefore leptons
within an angle of $|\cos \theta| < 0.95$ around the beam line and jets with
$|\cos \theta| < 0.90$ are discarded. After these numerical adjustments
the remaining isolated jets and leptons define the signature of the simulated
event.

Background from two-photon interactions was not simulated for this study, since
they typically lead to very soft hadronic events, and can be removed by a cut on
the total transverse momentum, $p_{\rm t}^{\rm tot} > 12$ GeV \cite{stop}.

The other backgrounds are reduced by the following simple kinematic cuts:
\begin{itemize}
\item Each event must contain four hadronic jets and no isolated
lepton and $p_{\rm t}^{\rm tot} > 12$ GeV (see above).
\item For each combination of two jets, the invariant mass is required to be
different from the $Z$ mass, $|m_{jj} - \MZ| > 10$ GeV. This strongly reduces
background from gauge bosons and neutralinos.
\item Since backgrounds from gauge bosons tend to increase in the forward and
backward detector regions, they can be further reduced by a cut on total
momentum angle, $|\cos \theta_{p_{\rm tot}}| = |p_{\rm long,tot}/p_{\rm tot}| <
0.9$.
\item All Standard Model backgrounds are cut down by requiring large missing
energy, $\Eslash > 100$ GeV.
\item Since two bottom jets are expected in the signal, the signal-to-background
ratio is improved through b-tagging. Following \cite{higgspair}, it is assumed
that
the b tagging efficiency is 90\%, with a mistag probability of light flavors of
10\%. B-tagging works on all backgrounds except $t\bar{t}$. For the other
backgrounds the four jets mainly originate from $W$ decays, so that two of the
jets could be charmed. Charm jets have a higher b-mistagging probability than
light flavors, but since the signal also always contains two charm jets, the
resulting tagging power between signal and gauge boson backgrounds is governed
by the  discrimination between bottom and light flavors.
\end{itemize}
With this selection procedure, the remaining background is very small, about
3 fb, whereas
the total signal efficiency is 18\%. Including an overall systematic acceptance
of 90\%, the resulting signal efficiency is 16\%. With a total luminosity of 250
fb$^{-1}$ the statistical error for the
cross-section measurements is
\begin{equation}
\delta\sigma_{\rm RL}[\chi^+_1\cha^-_1] = 0.37\%, \qquad
\delta\sigma_{\rm LR}[\chi^+_1\cha^-_1] = 1.6\%,
\label{eq:chaxsec}
\end{equation}
where RL and LR stand for the polarization combinations $P(e^+)$/$P(e^-)$ =
$+50$\%/$-80$\% and $-50$\%/$+80$\%, respectively.

The spectrum of the decay products can be used for a determination of the
chargino and neutralino mass. The spectrum of the invariant mass of the b and c
jet from the decay $\cha_1^+ \to \tilde{t} \bar{b} \to c \bar{b} \neu_1$ has a
characteristic upper edge at
\begin{equation}
m_{\rm \bar{b}c,max} = m_{\rm b\bar{c},max} = \mcha{1} - \mneu{1}.
\end{equation}
However, with a four jet final state the problem remains to identify the pairs
of two jets each that belong to the decay of one chargino. The bottom jets can
be identified as the jets with the highest b tagging likelihood. However this
still leaves a twofold ambiguity to combine a b jet with a c jet. Here the
following strategy is adopted: since a pair of jets originating from different
charginos tend to have a larger invariant mass than a pair originating from the
same chargino, always the pair with the lower invariant mass is selected for
the mass measurement.

\begin{figure}[tb]
\centering
\epsfig{file=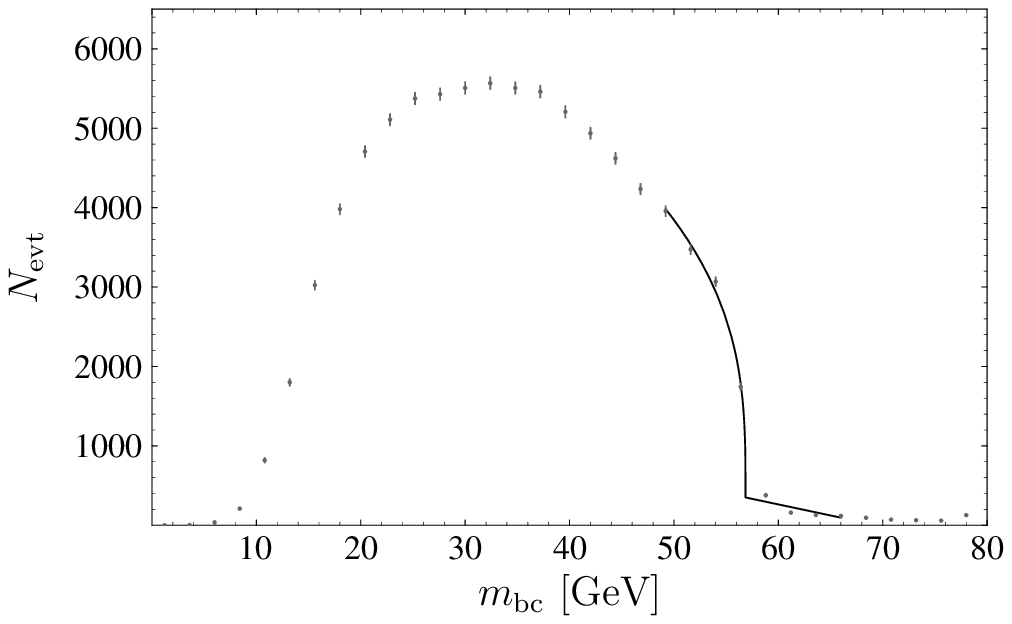, width=12cm}
\mycaption{Distribution of the bottom-charm invariant mass $m_{\rm bc}$ from
chargino decay and a simple fit to the upper edge.}
\label{fig:chadist}
\end{figure}
The resulting invariant mass distribution in shown in Fig.~\ref{fig:chadist}. A
very crude fit of the upper threshold gives
\begin{equation}
m_{\rm bc,max} = (56.6^{+0.22}_{-0.06} \pm 0.24) \gev, \label{eq:chadist}
\end{equation}
compared to the model input value $\mcha{1} - \mneu{1} = 56.8$ GeV. Here the
first error is the statistical uncertainty of the fit, whereas the second error
indicates the systematic uncertainty stemming from the jet energy scale. Based
on studies at LEP for $W$ boson pair production \cite{wwopal}, the jet energy
scale error is assumed to be 0.4\%. 

\begin{figure}[tb]
\centering
\epsfig{file=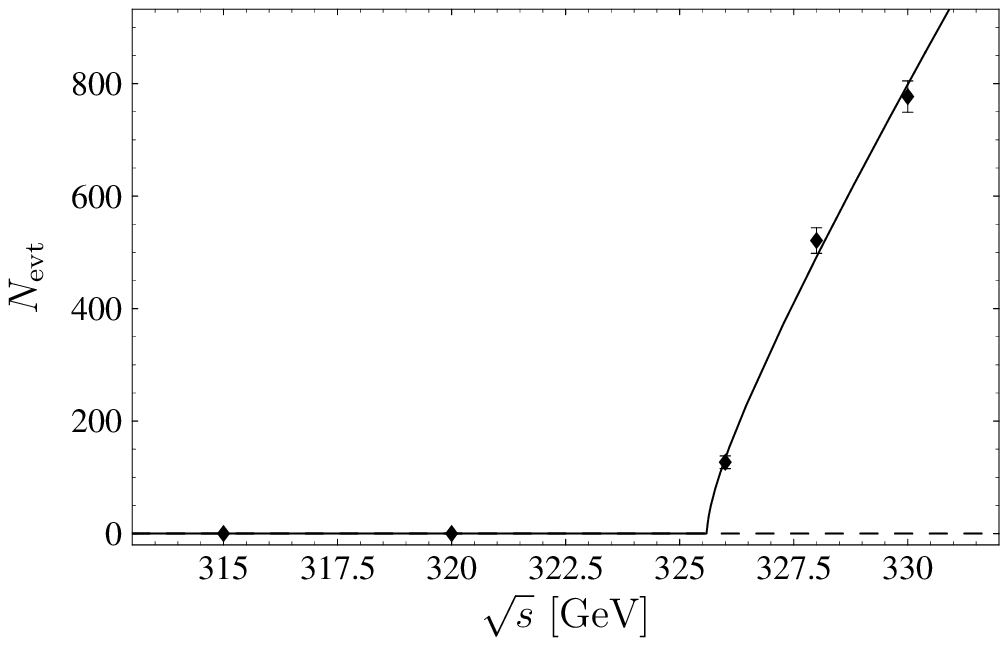, width=12cm}
\mycaption{Threshold scan for chargino pair production using five scan points
with 10 fb$^{-1}$ each.}
\label{fig:chathr}
\end{figure}

Since this measurement would only yield the difference between chargino and
neutralino mass, an independent measurement is necessary for the determination
of absolute values of both masses. A very precise direct determination of the
chargino mass can be obtained from a threshold scan. By measuring the chargino
cross-section at a few center-of-mass energies near the pair production
threshold, the onset of the pair production excitation curve can be
reconstructed and used for a mass determination. By including two points below
the threshold, the background can be analyzed and extrapolated in a
model-independent way. The result of the analysis for five scan points with
10 fb$^{-1}$ luminosity each, using the same methods and
selection cuts as above, is depicted in Fig.~\ref{fig:chathr}.
The result of a fit to the threshold excitation curve is
\begin{equation}
\mcha{1} = \tfrac{1}{2} \left[ \sqrt{s} \right]_{\rm thr} = 
	(163.02 \pm 0.04 \pm 0.04) \gev, \label{eq:chathr}
\end{equation}
where the first error is statistical and the second is the systematic error from
the beam energy calibration. Combining eqs.~\eqref{eq:chadist} and
\eqref{eq:chathr}, the masses are obtained as follows,
\begin{equation}
\mcha{1} = (163.02 \pm 0.06) \gev, \qquad
\mneu{1} = (106.1 \pm 0.3). \label{eq:chamass}
\end{equation}

\paragraph{Neutralinos \boldmath $\neu_2$ and $\neu_3$}
\anc\\
Neutralino production can have sizeable cross-section for mixed pair production
of $\neu_1\neu_3$ and $\neu_2\neu_3$. The signal for $\neu_1\neu_3$ is
characterized by two leptons or two jets plus missing energy, which receives
very large backgrounds from $W^+W^-$ and $ZZ$ production. Even after reducing
the backgrounds with appropriate cuts, the remaining level is still problematic
for any precision measurement. Therefore, in the following only $\neu_2\neu_3$
is considered.

This process has a sizeable cross-section, $\sigma[e^+e^- \to \neu_2
\neu_3] = 62$ fb for unpolarized beams. Using right-handed polarization for the
$e^-$ and left-handed polarization for the
$e^+$ beam, $P(e^+)$/$P(e^-)$ =
$-50$\%/$+80$\%, the signal is enhanced to 68.9 fb, while all Standard Model
backgrounds, which are dominated by left-chiral SU(2) interactions, 
are reduced for this beam polarization combination.

The $\neu_3$ almost always decays into a $Z$ boson and lightest neutralino
$\neu_1$, while the $\neu_2$, due to the small mass difference
$\mneu{2}-\mneu{1}$ decays through a virtual $Z$ into two quarks or two leptons.
In order to improve the statistical significance of the signal, the
hadronic decay modes of the $Z$ are considered, since the branching ratio of $Z$
to leptons is very small. Then the final state is characterized by four (light
quark) jets and missing energy.
Standard Model backgrounds arise from processes with two and three vector bosons
and from $t\bar{t}$ production. In addition, chargino production as discussed
above, constitutes another difficult background.

Signal and background are simulated and analyzed as detailed above.
The following cuts have been applied to select the signal out of the background:
\begin{itemize}
\item Similar to the chargino analysis, a four jets signature and the same cuts
on $\cos \theta_{p_{\rm tot}}$ and $\Eslash$ are used.
\item The invariant mass of two jets has to be equal to the $Z$ boson mass,
$|m_{j_aj_b} - \MZ| < 10$ GeV, whereas the invariant mass of the other two jets has
to be smaller than $\MZ$, $\MZ - m_{j_cj_d} > 10$ GeV. All combinatoric pairings
of the four jets are tried for this purpose, and if one combination meets these
requirement, the event is kept.
\item The Background from charginos generates rather soft jets due to the small
stop mass. Thus a cut on the transverse momentum, $p_{\rm t}^{\rm tot} > 50$ GeV
is
effective against this background.
\item Finally the chargino background is further reduced by a b-quark veto.
As a consequence, here the signal is restricted to light quark decay channels of
the $Z$ bosons only.
\end{itemize}
After this selection procedure, the remaining background is around 0.2 fb, but
a good signal efficiency of 26\% is achieved. 
Including a general systematic acceptance
of 90\%, the resulting signal efficiency is 24\%. 
With a total luminosity of 250
fb$^{-1}$ for the polarization combination $P(e^+)$/$P(e^-)$ = $-50$\%/$+80$\%,
the statistical error for the cross-section measurement is
\begin{equation}
\delta\sigma_{\rm RL}[\neu_2\neu_3] = 1.6\%.
\label{eq:neuxsec}
\end{equation}
For the opposite polarization combination $P(e^+)$/$P(e^-)$ = $-50$\%/$+80$\%,
the precision of the neutralino cross-section measurement is much worse due to
larger backgrounds, and does not improve the global MSSM analysis.

Information about the neutralino masses can be extracted from the decay product
distributions. In the $\neu_3 \to Z \neu_1$ decay, the energy spectrum of the
$Z$ is rather flat, with distinct lower and upper endpoints at
\begin{equation}
\begin{aligned}
E_{\rm min,max,3} &= \frac{1}{4\mneu{3}^2\sqrt{s}} 
\Bigl ( \mneu{3}^4 - \mneu{3}^2\mneu{2}^2 + \mneu{3}^2\MZ^2 -
 \mneu{2}^2\MZ^2 + \mneu{3}^2 s + \MZ^2 s \\ & \qquad\qquad\qquad - 
 \mneu{1}^2 (\mneu{3}^2 -
 \mneu{2}^2 + s) \mp \sqrt{\lambda(\mneu{3}^2, \mneu{1}^2, \MZ^2) \,
   \lambda(\mneu{3}^2, \mneu{2}^2, s)}\Bigr ),
\end{aligned}
\end{equation}
with $\lambda(a,b,c) = a^2 + b^2 + c^2 - 2 a b - 2 a c - 2 b c$. The energy of
the $Z$ boson can be deduced from the energy of the jet pair that combines to
the $Z$ invariant mass.

\begin{figure}[tb]
\begin{tabular}{p{8cm}p{8cm}} 
(a) \newline
\epsfig{file=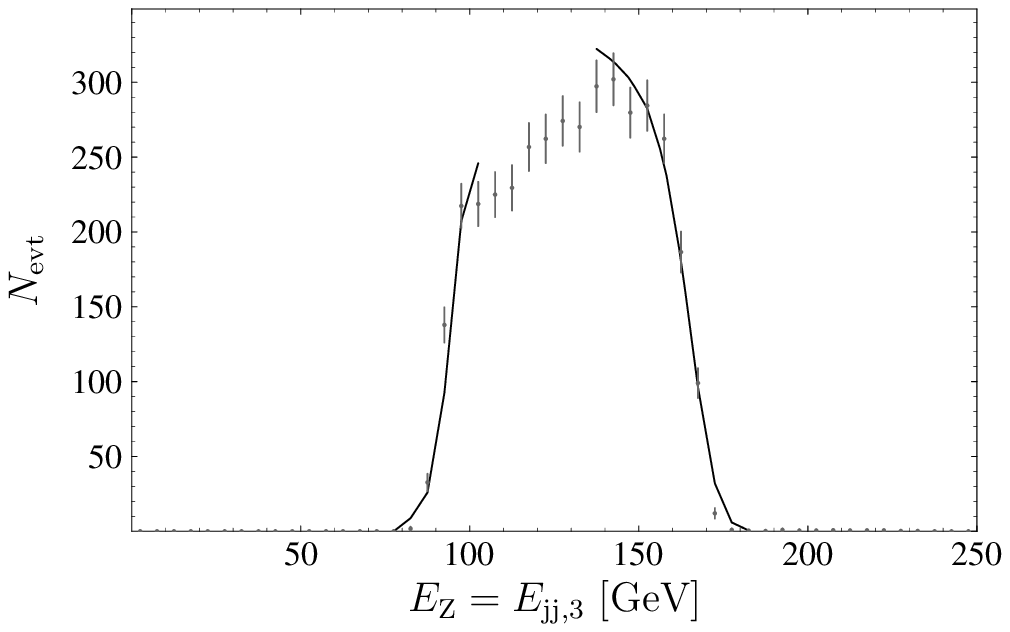, width=8cm} &
(b) \newline
\epsfig{file=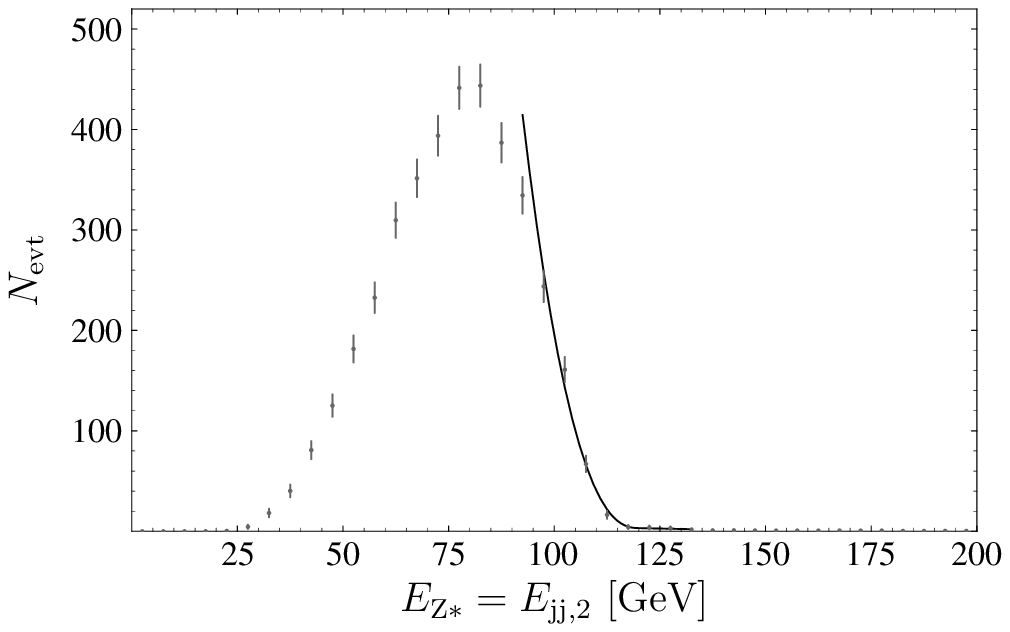, width=8cm} \\ 
(c) \newline
\epsfig{file=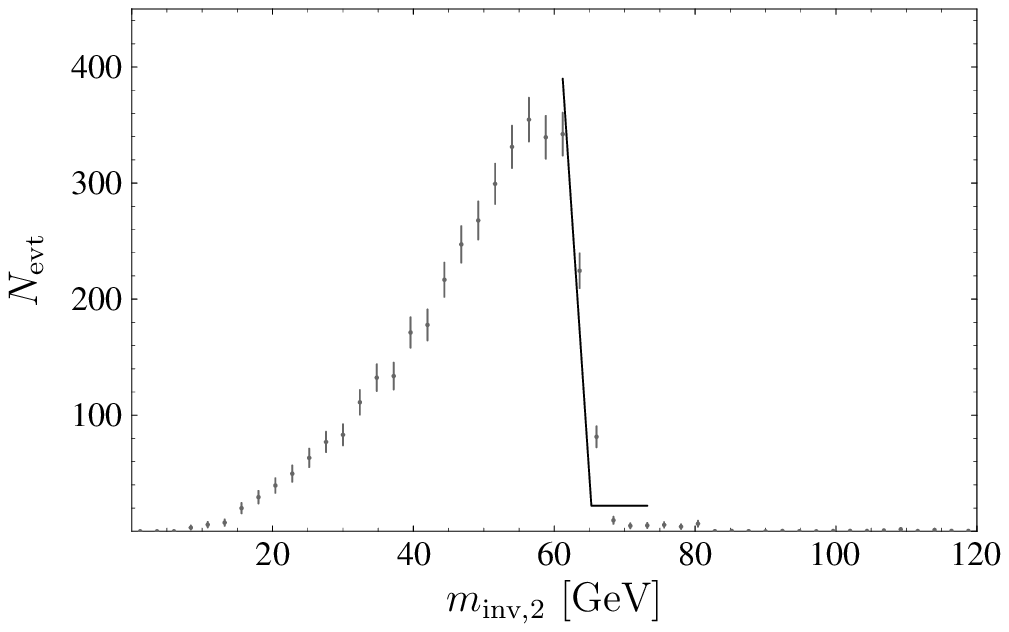, width=8cm} &
\anc\newline 
\anc\newline
\anc\newline
\mycaption{(a) Energy distribution for jet pair in $\neu_3 \to Z \neu_1$ decay,
(b) energy distribution for jet pair in $\neu_2 \to Z^* \neu_1$ decay,
and (c) invariant mass of the latter, including simple fitting
curves.\label{fig:neudist}} 
\end{tabular}
\end{figure}

For the $\neu_2$ decay, the kinematics are different since only a virtual $Z$
is involved in the process. Still the energy spectrum of the two final state
jets has a distinct upper endpoint given by
\begin{equation}
E_{\rm jj,max,2} = \frac{\mneu{2}^2 - \mneu{3}^2 - 2 \mneu{1}^2 \sqrt{s} +
s}{2\sqrt{s}}.
\end{equation}
In addition, the invariant mass of the same two jets is bounded from above by
the mass difference of the $\neu_2$ and $\neu_1$,
\begin{equation}
m_{\rm jj,max,2} = \mneu{2} - \mneu{1}.
\end{equation}
The simulated distributions after cuts and fits to the endpoints are summarized
in Fig.~\ref{fig:neudist}.
From a combination of the different fits, and including the jet energy scale error
as for the charginos, one obtains
\begin{equation}
\delta \mneu{1} = 1.3 \gev, \quad \delta \mneu{2} = 1.5 \gev, \quad
\delta \mneu{3} \approx 3.3 \gev.
\end{equation}
By feeding in the more precise determination of the lightest neutralino mass in
eq.~\eqref{eq:chamass}, these numbers are improved to
\begin{equation}
\delta \mneu{2} = 0.6\gev, \quad
\delta \mneu{3} \approx 2.0 \gev.
\label{eq:neumass}
\end{equation}

The mass measurements are not sufficient to extract the fundamental
supersymmetry parameters without ambiguity. Therefore the
additional information from measurements of the chargino and
neutralino cross-sections are important. The MSSM model parameters can be
extracted from a combined fit to all observables.

As a result of very heavy selectrons and sneutrinos, the t-channel contribution
to the production processes are essentially switched off. However, the
non-observation of these particles at colliders puts only a relatively mild
lower bound on their masses of about 500 GeV.
Here the values of the sleptons masses are kept
arbritrary, and instead indirect bounds on the masses are derived from the
fit (see also Ref.~\cite{FP2}).

The chargino and neutralino cross-section and mass measurements are combined in
a $\chi^2$ fit, thus taking into account all correlations in the underlying
parameters that enter in the observables. The fit results with
one standard deviation errors are
\begin{equation}
\begin{aligned}
M_1 &= (118.8 \pm 0.4) \gev, & \qquad \tan\beta &= 5^{+0.5}_{-0.7}, \\
M_2 &= (225.0 \pm 0.9) \gev,  \\
|\mu| &= (225.0 \pm 1.2) \gev, & \msn{{\rm e}} &> 5 \tev, \\
 |\phi_\mu| &< 0.6, & \mseR &> 1.5 \tev.
\label{eq:inobgen}
\end{aligned}
\end{equation}
The resulting constraint on the phase $\phi_\mu$ is rather poor, since none of
the included observables is directly CP-sensitive, and $\phi_\mu$ is strongly
correlated with other parameters, in particular $\tan\beta$.

\paragraph{CP violation}
\anc\\
The effect of CP violation can be uniquely studied in observables that are
directly CP-sensitive. In the chargino and neutralino sector, triple products of
kinematic momenta have been identified as useful for that purpose
\cite{triple,trip2,trip3}.

In this work, hadronic decays of neutralinos and charginos are investigated,
\begin{equation}
\begin{aligned}
e^+e^- \to \cha^s_i &\cha^{-s}_j && \qquad s = 0,\pm. \\
&\knickpfeil j_{\rm a} j_{\rm b} \neu_1
\end{aligned}
\end{equation}
From the momenta $p_{\rm a,b}$ of the two jets one can construct the T-odd
triple product 
\begin{equation}
{\cal T} = \vec{p}_{\rm e^-} (\vec{p}_{\rm a} \times \vec{p}_{\rm b}).
\end{equation}
and the T-odd asymmetry
\begin{equation}
{\cal A} = \frac{\sigma[{\cal T} > 0] - \sigma[{\cal T} < 0]}
		{\sigma[{\cal T} > 0] + \sigma[{\cal T} < 0]}.
\end{equation}
The experimental investigation of this triple product requires the
identification of the charge of the jets. While this is not possible for
light-quark jets on a event-by-event basis, it can be achieved for larger
samples of jet events on a statistical basis, see e.g. Ref.~\cite{l3jettag}.

Using the same Monte-Carlo techniques as explained above, the asymmetry is
calculated both for chargino and neutralino production. In the given scenario,
it is found that ${\cal A}[\cha^+_1\cha^-_1] = 0.3$\% and 
${\cal A}[\neu_2\neu_3] = 0.9$\%. Thus in both cases the expected magnitude is
of the same order as the statistical error of the cross-section measurements,
see eqs.~\eqref{eq:chaxsec}, \eqref{eq:neuxsec}.

For larger values of $\phi_\mu$, larger triple product asymmetries can be
obtained, however such models are already highly constrained by electric dipole
moment limits. Therefore a measurement of the CP-violating phase in the $\mu$
parameter appears to be hopeless at the ILC. These findings are in agreement
with previous theoretical studies \cite{trip2}.


\subsection{Cosmological implications}
\label{sec:dm}

The discovery of light scalar top quarks, in conjunction with a
Standard-Model-like Higgs boson with a mass near 120 GeV,  would be 
a strong indication that electroweak baryogenesis is
the mechanism for the generation of the baryon asymmetry.
At the same time, supersymmetry could also 
explain the existence of dark matter in the universe, based
on the co-annihilation mechanism. In order to confirm this exciting picture,
the relevant supersymmetry parameters have to be measured accurately.

One needs to (i) determine that the light stop is mainly right-chiral to
contribute appropriately to the electroweak phase transition while being in
agreement with electroweak precision measurements, (ii) check that the masses
and compositions of the gauge/Higgs superfield sector are compatible with the
values required for the generation of the  baryon asymmetry, and (iii) compute
the dark matter annihilation cross-sections and the relic abundance so as to
compare with cosmological observations. If stop-neutralino co-annihilation is
relevant it is important to determine the stop-neutralino mass difference very
precisely. 


\paragraph{Baryogenesis}
\anc\\
In the given scenario, the first information about the necessary constituents
for baryogenesis in the MSSM can be obtained at the LHC, namely the discovery
of a light Higgs boson and  possibly evidence for a light stop. At the
ILC, the picture can be rendered much more precisely by determining the mass and
mixing angle of the light stop accurately. 

Using the
computations of Refs.~\cite{CQW,Carena:2002ss} for the electroweak phase
transition, it is found that the experimental uncertainty of the stop
parameters allows to determine the strength of the phase transition to better
than 10\%,
\begin{equation}
\delta_{\rm exp}\left[\frac{v(T_{\rm c})}{T_{\rm c}}\right] \lesim 10\%.
\end{equation}
Note however that this is only a parametric error, and there are also
theoretical uncertainties involved in the computation.

Besides the strength of the electroweak phase transition, the second crucial
ingredient for electroweak baryogenesis is the generation of the baryon
asymmetry through CP-violating processes. As explained in section~\ref{sec:not},
in the MSSM these CP-violating contributions can be described by a complex
phase of the parameter $\mu$. However, the findings of the previous sections
show that for typical values of $\phi_\mu$, the CP-violating effects in
neutralino and chargino observables at the ILC are
too small to be observed. It would only be possible to derive an upper bound of
$|\phi_\mu| \lesim 0.7$ at the 90\% confidence level, which would
leave  the question of the baryon asymmetry still undecided.


\paragraph{Dark matter} \anc\\ As discussed widely in the literature (see
e.g.~\cite{stop,nojiri,peskin}),  collider experiment data can be used to
compute the  expected cosmological dark matter relic density. In the scenario
studied here, the relic density is governed by the neutralino-stop
co-annihilation mechanism, and thus crucially depends on the stop-neutralino
mass difference. It was found \cite{stopmass} that this mass difference cannot
be extracted with good precision directly from stop decay distributions,
because of radiation effects and limited statistics, but the best determination
is derived from independent mass measurements of the stop and neutralino. While
the stop mass and mixings can be extracted from precision measurements of the
stop cross-section \cite{stop}, the neutralino properties have to be determined
from independent observables as discussed in this report.

The relic dark matter density is computed with the codes described in
Ref.~\cite{Balazs:2004bu,morr}. The analysis is based on observables in the stop
and neutralino/chargino sectors. For the neutralinos and charginos, 
the estimated errors taken from the previous chapter, see
eqs.~\eqref{eq:chaxsec}, \eqref{eq:chamass}, \eqref{eq:neuxsec},
\eqref{eq:neumass}. The stop measurement errors are derived from the study of
Ref.~\cite{stop} with the result: $\delta \mst = 1.2$ GeV and
$|\cos\theta_{\tilde{t}}| < 0.077$.

In total, the relevant parameters used as input are $\mst$, $\cos
\theta_{\tilde{t}}$,  $M_1$, $M_2$, $|\mu|$, $\phi_\mu$, $\tan\beta$. It has
been checked that the dependence on Higgs and slepton parameters is negligible
for this scenario. The mass of the heavier stop $\tilde{t}_2$ is too large to
be measured directly, but it is supposed that a limit of $m_{\tilde{t}_2} >
1000$ GeV can be set from LHC searches.

\begin{figure}[tb]
\centering
\epsfig{file=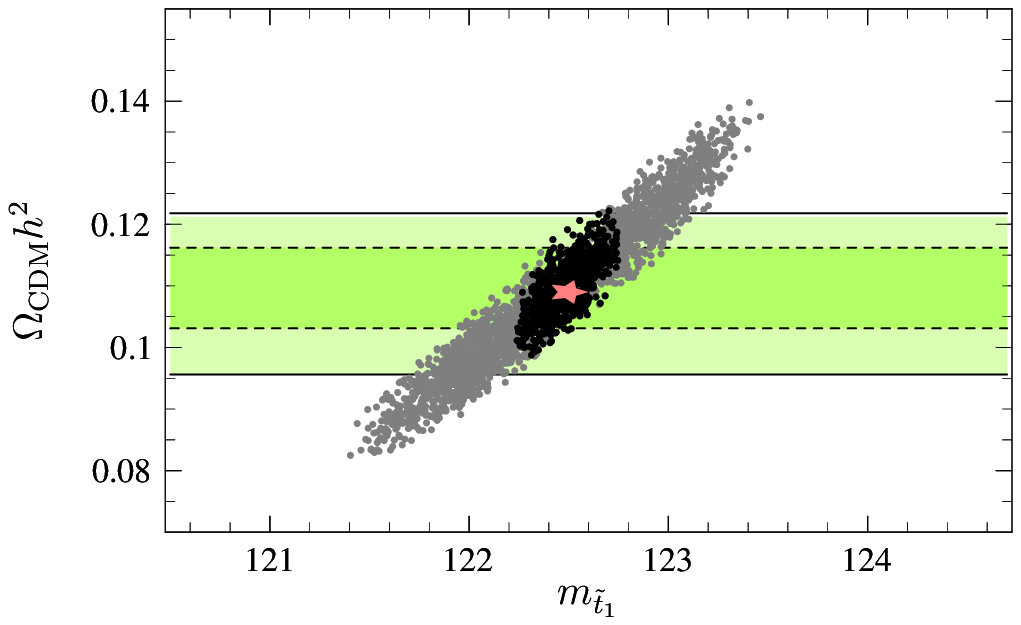, width=13cm}
\mycaption{Computation of dark matter relic abundance $\Omega_{\rm CDM} h^2$ for the
scenario BGEN,
taking into account estimated experimental errors for stop, chargino, neutralino
and Higgs sector measurements at future colliders. The dots correspond to
a scan over the 1$\sigma$ ($\Delta \chi^2 \leq 1$) region allowed by the 
experimental errors, as a
function of the measured stop mass, for a mass measurement error of 1.2 GeV
(gray dots) and 0.3 GeV (black dots). The original scenario used as input is
indicated by the red (light gray) star. The horizontal shaded bands show the
1$\sigma$ and 2$\sigma$ constraints on the relic
density measured by WMAP and SDSS.}
\label{fig:dm}
\end{figure}
The expected experimental errors are propagated and parametric correlations are
taken into account by means of a $\chi^2$ scan. However, theoretical errors due
to missing  radiative corrections are not taken into account in the calculation
of the dark matter density. Fig.~\ref{fig:dm} shows the result of a scan over
the MSSM parameter space for the scenario BGEN.  The scattered gray dots
indicate the region allowed by the collider experimental uncertainty, as a
function of the measured stop mass. The range of the horizontal axis is
constrained by the error in the stop mass measurement, $m_{\tilde{t}_1} =
(122.5 \pm 1.2) \gev$. The horizontal bands depict the relic density as
measured by WMAP and SDSS \cite{wmap,sdss} 
with one and two standard deviation errors. At
1$\sigma$ level, the astrophysical observations lead to $0.103 < \Omega_{\rm
CDM} h^2 < 0.116$.

In total, using the collider measurement simulations, the relic density can be
predicted to $0.082 < \Omega_{\rm CDM} h^2 < 0.139$ at the 1$\sigma$ level.
Thus the overall precision is of the same magnitude as, but worse by roughly a
factor 4 than the direct WMAP/SDSS determination. The uncertainty in the
theoretical
determination is dominated by the measurement of the $\tilde{t}_1$ mass with an
error of 1.2 GeV. This shows that even in a scenario with heavy sfermions,
which is unfavorable for future collider experiments, the achievable precision
in the analysis of the chargino and neutralino sector is very high and not a
limiting factor for cosmological interpretations.

First results of an optimized threshold scan method indicate that the
precision for  $\mst$ can be improved to about 0.3 GeV~\cite{mschmitt}. The
advantage of the threshold scan method is the small influence of systematic
errors, since it makes use of the shape of the cross-section as a function of
the center-of-mass energy, instead of absolute cross-section measurements.
However it is limited by small statistics near the threshold. The best accuracy
can be achieved by combining a measurement near the threshold, where the
cross-section is most sensitive to the stop mass, with a measurement at higher
energies, where the cross-section is larger~\cite{mschmitt}.
With a stop mass error of $\delta \mst = 0.3 \gev$, the relic density could be
computed much more precisely, yielding the result $0.099 < \Omega_{\rm CDM} h^2
< 0.122$ in the scenario BGEN. This precision is very
comparable to the direct WMAP/SDSS determination, as indicated by the black dots in
Fig.~\ref{fig:dm}.


\section{Focus-point scenario}
\label{sec:fp}

\subsection{Chargino and neutralino studies at the LHC}

In a focus-point scenario, all sfermions are very heavy, but gluinos
$\tilde{g}$ can be sufficiently light to be generated with large cross-sections.
The gluino decays can proceed via some neutralinos and charginos, giving a
clear and distinct leptonic discovery signature \cite{tadas}. For example, a
signature with same-sign leptons and missing transverse energy can be
interpreted as a decay cascade of the gluinos via charginos \cite{qcdyuk}.
Besides, the decay $\neu_2 \to Z^* \neu_1 \to l^+l^- \neu_1$ allows to
determine the mass difference $m_{21} = \mneu{2}-\mneu{1}$ from the $l^+l^-$
invariant mass distribution with a precision of about  $\delta m_{21} \sim (0.5
\gev) \times  \exp(-m_{\tilde{g}}/(200 \gev)+3.5)$, where the  exponential
factor with $m_{\tilde{g}}$ approximately accounts for the decrease of the
gluino cross-section with larger gluino masses.

\subsection{Chargino and neutralino studies at the ILC}

In the scenario LCC2, among the superpartners, only charginos and neutralinos
can be studied at a future linear collider, whereas all sfermions are too
massive to be accessible either at the LHC or ILC. This situation arises always
in focus point supersymmetry and has been studied in previous works \cite{FP,
FP2}. Here the study of Ref.~\cite{FP2} is extended by including different
observables and simulations for signal and background processes.

As in the previous section, the general procedure for reconstructing the
underlying MSSM parameters  relies on the assumption of the MSSM structure for
the chargino and neutralino mass  matrices. 
The parameters $M_1$, $M_2$, $|\mu|$, $\phi_\mu$ and $\tan\beta$ are extracted
from a fit to mass and cross-section measurements.

For the scenario LCC2 (see appendix), the chargino and neutralino masses at
tree-level amount to
\begin{equation}
\begin{aligned}
\mneu{1} &= 107.7 \gev, & \mneu{2} &= 166.3 \gev, & \mcha{1} &= 159.4 \gev, \\
\mneu{3} &= 190.0 \gev,	& \mneu{4} &= 294.4 \gev, & \mcha{2} &= 286.8 \gev. \\
\end{aligned}
\end{equation}
As pointed out above, the LHC has a good possibility to find
evidence for and measure some properties of neutralinos and charginos. At the
ILC with $\sqrt{s} = 500$ GeV, many neutralino and chargino states are
accessible, see Tab.~\ref{tab:ilcXsec2}.
\renewcommand{\arraystretch}{1.2}
\begin{table}[tp]
\begin{center}
\begin{tabular}{|r||rrr|}
\hline
$e^+e^- \to \neu_i \neu_j$ & $\neu_i = \neu_2$ & $\neu_3$ & $\neu_4$ \\
\hline \hline
$\neu_j = \neu_1$ & 8.0 & 46.3 & 0.01 \\
$\neu_2$ & 8.7 & 107.2 & 0.07 \\
$\neu_3$ && 6.2 & 22.7 \\
$\neu_4$ &&& --- \\
\hline \hline
$e^+e^- \to \cha^\pm_i \cha^\mp_j$ & $\cha^\pm_i = \cha^\pm_1$ & $\cha^\pm_2$ & \\
\hline \hline
$\cha^\mp_j = \cha^\mp_1$ & 534 & 23 & \\
$\cha^\mp_2$ & & --- & \\
\hline
\end{tabular}
\end{center}
\vspace{-1em}
\mycaption{Tree-level production cross-sections in fb at $\sqrt{s} = 500$ GeV
with unpolarized beams
for the reference point LCC2.}
\label{tab:ilcXsec2}
\end{table}
The most promising processes are $e^+e^- \to \cha^+_1 \cha^-_1$, $e^+e^- \to
\neu_1 \neu_3$ and $e^+e^- \to \neu_2 \neu_3$, similar to the baryogenesis
scenario.

Tab.~\ref{tab:decays2} gives the relevant decay modes and branching fractions.
\renewcommand{\arraystretch}{1.2}
\begin{table}[tb]
\begin{center}
\begin{tabular}{|c||c|c|r@{\:}lr|}
\hline
Sparticle & Mass $m$ [GeV] & Width $\Gamma$ [GeV]
          & \multicolumn{3}{c|}{Decay modes} \\
\hline \hline
$\neu_1$ & $107.7$ & --- & \multicolumn{2}{c}{---} & \\
$\neu_2$ & $166.3$  & $0.000016$ &
        $\neu_2$ & $\to Z^* \, \neu_1$ & 100\% \\
$\neu_3$ & $190.0$  & $0.0013$ &
        $\neu_3$ & $\to Z^* \, \neu_1$ & 100\% \\
        &&&& $\to h^{0*} \, \neu_1$ & $< 1$\% \\
$\neu_4$ & $294.4$  & $0.81$ &
        $\neu_4$ & $\to Z \, \neu_1$ & $\ll 1$\% \\
        &&&& $\to Z \, \neu_2$ & 1\% \\
        &&&& $\to Z \, \neu_3$ & 1\% \\
        &&&& $\to W^\pm \, \cha^\mp_1$ & 94\% \\
        &&&& $\to h^0 \, \neu_1$ & $\ll 1$\% \\
        &&&& $\to h^0 \, \neu_2$ & 3\% \\
\hline
$\cha_1^\pm$ & $159.4$  & $0.0002$ &
        $\cha^+_1$ & $\to W^{+*} \, \neu_1$ & 100\% \\
$\cha_2^\pm$ & $286.8$  & $0.73$ &
        $\cha^+_2$ & $\to W^+ \, \neu_1$ & 3\% \\
        &&&& $\to W^+ \, \neu_2$ & 42\% \\
        &&&& $\to W^+ \, \neu_3$ & 11\% \\
        &&&& $\to Z \, \cha^+_1$ & 36\% \\
        &&&& $\to h^0 \, \cha^+_1$ & 8\% \\
\hline
\end{tabular}
\end{center}
\vspace{-1em}
\mycaption{Tree-level masses, widths and main branching ratios of the
neutralino and chargino states at Born level
for the reference point LCC2.}
\label{tab:decays2}
\end{table}
In this scenario, the mass differences between the lighter neutralino and
chargino states  are rather small, thus allowing only decays of the light
states through virtual gauge  bosons.

\paragraph{Chargino \boldmath $\cha_1^+$}
\anc\\
Production of light charginos, $e^+e^- \to \cha_1^+ \cha_1^-$, has a large
cross-section of 534 fb at $\sqrt{s} = 500$ GeV, which can be further increased
to 1345 fb by using beam polarization, $P(e^+)$/$P(e^-)$ = $+50$\%/$-80$\%.

In the focus-point scenario, charginos almost always decay through virtual $W$
bosons, $e^+e^- \to \cha_1^+ \cha_1^- \to W^{+*} W^{-*} \, \neu_1 \, \neu_1 \to
l^\pm \nu_l \, q \bar{q}' \, \neu_1 \, \neu_1$ where it is useful to consider a
final state where one $W^*$ decays hadronically into quarks, while the other
decays into a lepton $l = e,\mu$ and a neutrino. This final state allows a
precise kinematical measurement, since the two intermediate virtual $W$ bosons
can be disentangled, as opposed to the purely hadronic final state. As before,
Standard Model background from vector bosons and $t\bar{t}$, as well as
supersymmetric background from neutralino production is taken into account.

Signal and background events are simulated with the same Monte-Carlo methods as
elaborated above. The following selection cuts are applied to reduce the
backgrounds:
\begin{itemize}

\item Each event must contain two hadronic jets, one isolated
lepton and $p_{\rm t}^{\rm tot} > 12$ GeV.

\item The invariant mass of the two jets is required to be smaller than the $W$
mass, $\MW - m_{jj} < 10$ GeV, to reduce Standard Model background.

\item All Standard Model backgrounds are cut down by requiring large missing
energy, $\Eslash > 100$ GeV, and the direction of the missing momentum to be in
the visible detector region, $|\cos \theta_{p_{\rm miss}}| < 0.8$.

\item Since a large background comes from $W^+W^-$ production, the invariant
mass of the lepton and the missing momentum, $m_{lp_{\rm miss}}$ tends to
increase near the $W$ mass. Thus by requiring $m_{lp_{\rm miss}} > 150$ GeV,
that background is reduced.

\end{itemize}
After these cuts, the remaining background is very small, but 39\% of the
signal is retained. With an overall systematic acceptance of 90\% and a total
luminosity of 250 fb$^{-1}$ the statistical error for the cross-section
measurements is
\begin{equation}
\delta\sigma_{\rm RL}[\chi^+_1\cha^-_1] = 0.6\%, \qquad
\delta\sigma_{\rm LR}[\chi^+_1\cha^-_1] = 1.7\%,
\label{eq:chaxsec2}
\end{equation}
where RL and LR stand for the polarization combinations $P(e^+)$/$P(e^-)$ =
$+50$\%/$-80$\% and $-50$\%/$+80$\%, respectively.

The $\cha_1^\pm$ and $\neu_1$ masses can be constrained from distributions of
the decay products. The summed energy of the two jets from the decay
$\cha^\pm_1 \to W^\pm \neu_1 \to q\bar{q}' \neu_1$ have a well-defined upper
endpoint at
\begin{equation}
E_{\rm jj,max} = \frac{\sqrt{s}}{4} \left(1-\frac{\mneu{1}^2}{\mcha{1}^2}
\right)
  \left(1+ \sqrt{1-4\mcha{1}^2/s} \right).
\end{equation}
Moreover, the invariant mass of the same two jets is bounded from above by
\begin{equation}
m_{\rm jj,max}  = \mcha{1}-\mneu{1}.
\end{equation}
Simple fits to the upper tails of the energy and invariant mass distributions
give
\begin{equation}
E_{\rm jj,max} = (120.1 \pm 0.6 \pm 0.5) \gev, \qquad
m_{\rm jj,max} = (51.8 \pm 0.14 \pm 0.2) \gev, \label{eq:chadist2}
\end{equation}
compared to the model input values $E_{\rm jj,max} = 120.2$ GeV and $\mcha{1} -
\mneu{1} = 51.7$ GeV. Here the first error is the statistical uncertainty of
the fit, whereas the second error accounts for the 0.4\% jet energy scale
uncertainty. These measurement translate into the following results for the
masses:
\begin{equation}
\mcha{1} = (159.4 \pm 1.0) \gev, \qquad
\mneu{1} = (107.7 \pm 0.9) \gev,
\end{equation}
with a large correlation between the two masses. With an independent
determination of one of the masses from a different observable, both mass
determinations can be improved substantially.

As before, a threshold scan of the chargino cross-section yields a much more
precise determination of the chargino mass. Using five scan points with 10
fb$^{-1}$ luminosity each and the same  selection cuts as above, a fit to the 
threshold excitation curve gives
\begin{equation}
\mcha{1} = \tfrac{1}{2} \left[ \sqrt{s} \right]_{\rm thr} = 
	(159.38 \pm 0.04 \pm 0.04) \gev, \label{eq:chathr2}
\end{equation}
where the first error is statistical and the second is the systematic error
from the beam energy calibration. Combining eqs.~\eqref{eq:chadist2} and
\eqref{eq:chathr2}, the masses are obtained as follows,
\begin{equation}
\mcha{1} = (159.38 \pm 0.06) \gev, \qquad
\mneu{1} = (107.7 \pm 0.21). \label{eq:chamass2}
\end{equation}

\paragraph{Neutralinos \boldmath $\neu_2$ and $\neu_3$}
\anc\\
As for the baryogenesis scenario, only the neutralino production process
$\neu_2\neu_3$ yields a good signal-to-background ratio. This process has a
sizeable cross-section, $\sigma[e^+e^- \to \neu_2 \neu_3] = 107$ fb for
unpolarized beams. Using right-handed polarization for the $e^-$ and
left-handed polarization for the $e^+$ beam, $P(e^+)$/$P(e^-)$ =
$-50$\%/$+80$\%, the signal is enhanced to 129 fb, while all Standard Model
backgrounds are reduced for this beam polarization combination.

Both the $\neu_2$ and $\neu_3$ almost always decay through a virtual $Z$ boson
to a pair  of leptons or quarks and the lightest neutralino. As a consequence,
the decay products of the two neutralinos cannot be distinguished by
kinematical constraints. Therefore, here the semi-leptonic decay channel is
investigated, with one $Z$ decaying leptonically, and the other $Z$ decaying
hadronically. Thus the final state is characterized by two charged leptons, two
jets and missing energy. As for the baryogenesis case, the relevant Standard
Model backgrounds arise from processes with two and three vector bosons and
from $t\bar{t}$ production. 

\begin{figure}[tb]
\begin{tabular}{p{8cm}p{8cm}} 
(a) \newline
\epsfig{file=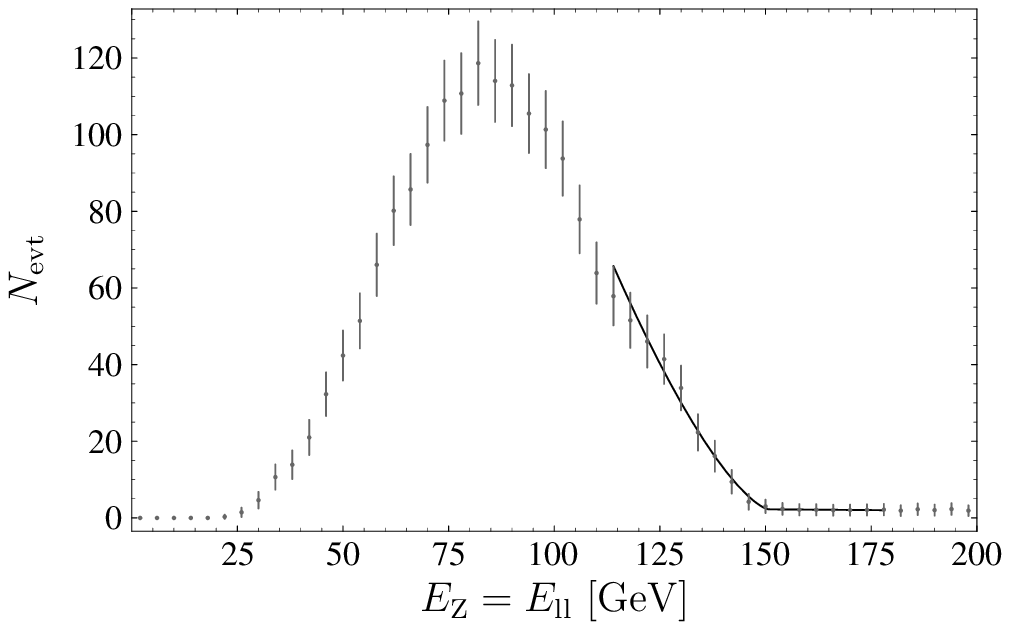, width=8cm} &
(b) \newline
\epsfig{file=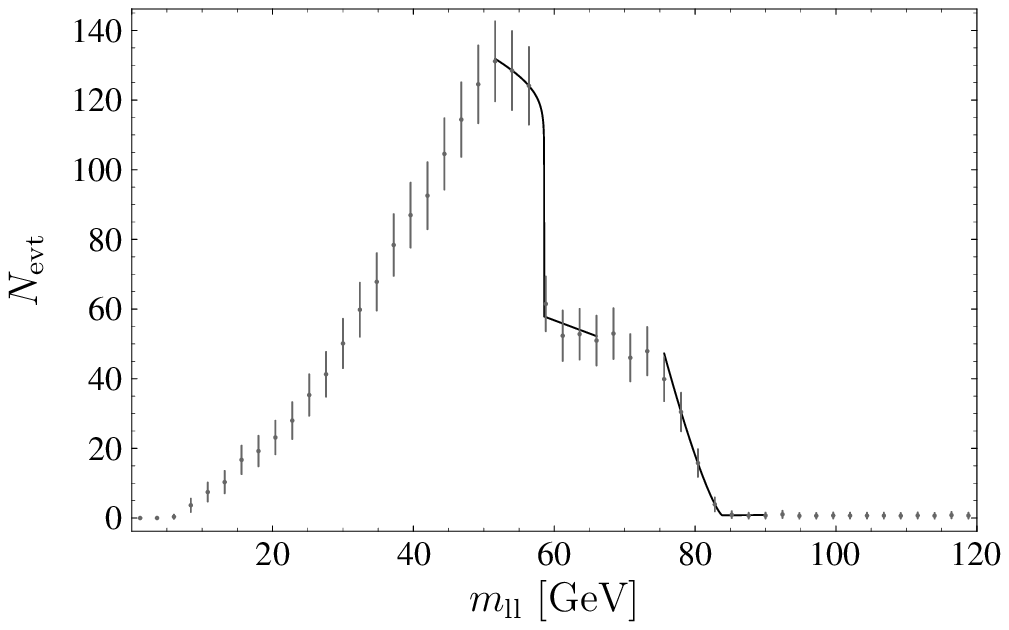, width=8cm}
\end{tabular}
\mycaption{(a) Energy distribution for lepton pair in $\neu_{2,3} \to Z \neu_1$ decays,
(b)  invariant mass of the latter, including simple fitting curves.} 
\label{fig:neudist2}
\end{figure}

The selection cuts to extract the signal from the background are very similar
to the baryogenesis case:
\begin{itemize}
\item Each event must contain two hadronic jets and two isolated
lepton and $p_{\rm t}^{\rm tot} > 12$ GeV.
\item The invariant mass of either the two jets or the two leptons  has
to be smaller than $\MZ$, $\MZ - m_{jj} > 10$ GeV and $\MZ - m_{ll} > 10$ GeV. 
\item As in section~\ref{sec:ilc}, it is required that
 $|cos \theta_{p_{\rm tot}}| < 0.9$ and $\Eslash > 100$ GeV and
a b-quark veto again $t\bar{t}$ background is applied.
\end{itemize}
A resulting signal efficiency of 47\% is obtained. 
With a total luminosity of 250 fb$^{-1}$ for the
polarization combination $P(e^+)$/$P(e^-)$ = $-50$\%/$+80$\%, the statistical
error for the cross-section measurement is
\begin{equation}
\delta\sigma_{\rm RL}[\neu_2\neu_3] = 2.7\%.
\label{eq:neuxsec2}
\end{equation}
For measurements of kinematic distributions, it is useful
to restrict oneself to the two leptons in the final state only, which can be
measured more cleanly and
precisely than jets. The summed energy of the lepton pair has a distinct upper
endpoint, depending from which of the two neutralinos they originate,
\begin{align}
E_{\rm ll,max,2} &= \frac{\mneu{2}^2 - \mneu{3}^2 - 2 \mneu{1}^2 \sqrt{s} +
s}{2\sqrt{s}}, \\
E_{\rm ll,max,3} &= \frac{\mneu{3}^2 - \mneu{2}^2 - 2 \mneu{1}^2 \sqrt{s} +
s}{2\sqrt{s}}.
\end{align}
In addition, the invariant mass spectrum of the same two leptons has an
endpoint at the mass difference between the intermediate and the final state
neutralino,
\begin{equation}
m_{\rm ll,max,2} = \mneu{2} - \mneu{1}, \qquad m_{\rm ll,max,3} = \mneu{3} - \mneu{1}
\end{equation}
The simulated distributions after cuts and fits to the endpoints are summarized
in Fig.~\ref{fig:neudist2}.
For the masses in Tab.~\ref{tab:decays2}, the upper endpoint of the $E_{\rm ll,2}$ spectrum is smaller than the maximum of the $E_{\rm ll,3}$ spectrum originating from the $\neu_3$. As a result, the endpoint $E_{\rm ll,max,2}$ gets diluted in the continuum of the $E_{\rm ll,3}$ distribution. Therefore, here only the endpoint $E_{\rm ll,max,3}$ is fitted, but no numerical value for $E_{\rm ll,max,2}$ is obtained. The $m_{\rm ll}$ shows two characteristic edges corresponding to the contribution from the $\neu_2$ and the $\neu_3$.

Combining fits to these three endpoints, and including a lepton energy scale
error of 0.l\%, one obtains
\begin{equation}
\delta \mneu{1} = 2.9 \gev, \quad \delta \mneu{2} = ^{+2.6}_{-3.1} \gev, \quad
\delta \mneu{3} = ^{+4.8}_{-4.0} \gev.
\end{equation}
By feeding in the more precise determination of the lightest neutralino mass in
eq.~\eqref{eq:chamass2}, these numbers are improved to
\begin{equation}
\delta \mneu{2} = ^{+0.3}_{-2.0} \gev, \quad
\delta \mneu{3} = ^{+2.6}_{-2.3} \gev.
\label{eq:neumass2}
\end{equation}

From a fit to the cross-sections, not only constraints on the chargino and
neutralino parameters are obtained, but also indirect bounds are derived for the
masses of the heavy sleptons appearing in the t-channel production contribution
(see also Ref.~\cite{FP2}).

The result of the combined fit, taking into account chargino and neutralino
cross-section and mass measurements, is
\begin{equation}
\begin{aligned}
M_1 &= (123.1^{+0.4}_{-0.3}) \gev, & \qquad \tan\beta &= 10^{+0.8}_{-1.4}, \\
M_2 &= (237.6^{+0.7}_{-1.1}) \gev,  \\
|\mu| &= (178.6^{+0.5}_{-0.5}) \gev, & 2.6 \tev < \msn{{\rm e}} &< 4.8 \tev, \\
 |\phi_\mu| &< 0.6, & \mseR &> 1.2 \tev.
\label{eq:inofcp}
\end{aligned}
\end{equation}
where the errors indicate one standard deviation (1$\sigma$) uncertainties.
The precision for the underlying MSSM parameters is comparable to the findings
of Ref.~\cite{FP2}, although this comparison is precarious 
since the focus-point scenario studied there is different from the point LCC2.
The underlying scenario LCC2 does not contain any CP-violating phases, but
nevertheless a non-zero phase of the $\mu$ parameter was allowed in the fit.
It turns out that the mass and cross-section observables give only a rather poor
constraint on $\phi_\mu$.


\subsection{Cosmological implications: dark matter}
\label{sec:dm2}

In focus-point supersymmetry, the annihilation cross-section is driven by a
sizeable higgsino component of the lightest neutralino. As a result, the
neutralinos can annihilate efficiently into electroweak gauge bosons. In order
to scrutinize this scenario, the neutralino properties need to be determined
with high precision. For the sfermions, only indirect bounds can be obtained,
but this is sufficient to establish the fact that they effectively decouple from
the annihilation cross-section.

The relic dark matter density is computed with {\sc DarkSUSY~4.1}
\cite{darksusy}. Including the expected precision for the hypothetical
neutralino and chargino measurements discussed in the previous section, see
eqs.~\eqref{eq:chaxsec2}, \eqref{eq:chamass2}, \eqref{eq:neuxsec2},
\eqref{eq:neumass2}, the accuracy for the dark matter determination is obtained
from a $\chi^2$ scan. It is
assumed that a limit of $m_{\tilde{q}} > 1000$ GeV for the squark masses 
can be set from non-observation of these particles at the LHC. While in many
scenarios it is possible to explore much larger squark masses, a bound of 1000
GeV is sufficient to guarantee that the effect of the squarks on the dark matter
annihilation is negligible.

\begin{figure}[tb]
\centering
\epsfig{file=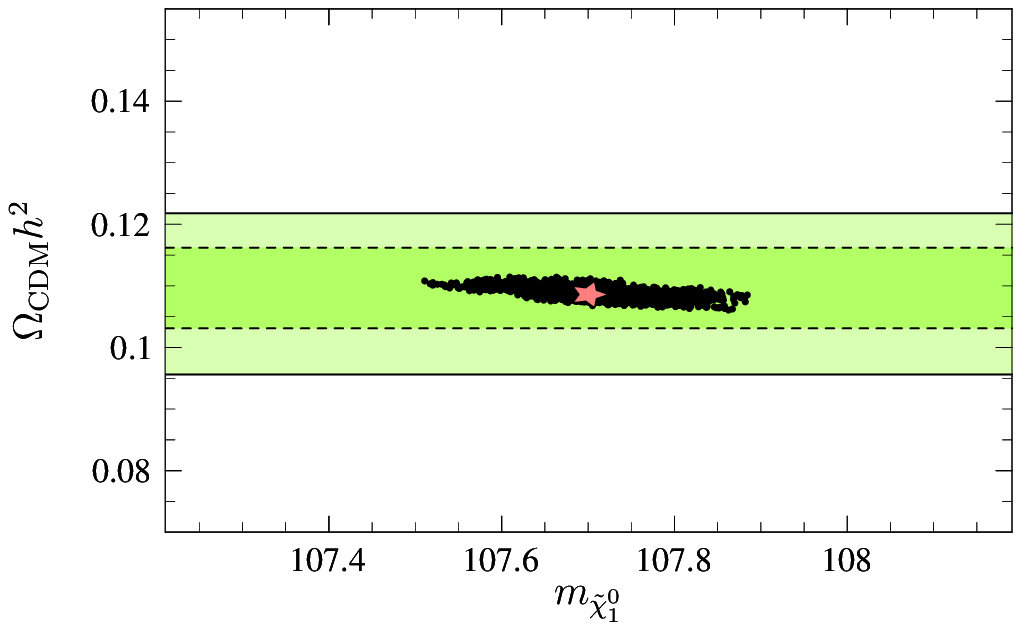, width=13cm}
\mycaption{Computation of dark matter relic abundance $\Omega_{\rm CDM} h^2$ for the
scenario LCC2, as a function of the 
lightest neutralino mass $\mneu{1}$.
The black dots are the 1$\sigma$ allowed region.
The original scenario used as input is
indicated by the red (light gray) star. The horizontal shaded bands show the
1$\sigma$ and 2$\sigma$ constraints on the relic
density measured by WMAP and SDSS.}
\label{fig:dm2}
\end{figure}
Fig.~\ref{fig:dm2} depicts the result of the scan for the scenario LCC2.  The
scattered black dots indicate the region allowed by the collider experimental
uncertainty, as a function of $\mneu{1}$. As pointed out above, although the
LCC2 scenario is CP-conserving, a CP-violating phase for $\mu$ was allowed in
the fit to the
mass and cross-section measurements. However it was checked that when
constraining $\phi_\mu$ to zero in the fit, the results remain essentially
unchanged. The reason for this is that the annihilation cross-section only
mildly depends on the CP-phase, but much more strongly on the neutralino mass
parameters,
which are directly measured. The precision of the predicted relic 
density $\Omega_{\rm CDM} h^2$ is a very remarkable 2.5\% at 1$\sigma$ level.

The computed relic density depends sensitively on the lightest neutralino mass
$\mneu{1}$.
Compared to
Refs.~\cite{FP,peskin}, the neutralino mass determination is improved in this
analysis by the inclusion of the chargino threshold scan. As a result, the
dark matter computation from the hypothetical collider measurements is also more
accurate, with $\delta[\Omega_{\rm CDM} h^2] \sim 2.5$\% instead of 
$\delta[\Omega_{\rm CDM} h^2] \sim  8$\% in Ref.~\cite{peskin}.


\section{Impact of positron polarization on ILC results}
\label{sec:pol}

Throughout the previous sections, is was assumed that both the $e^-$ and the
$e^+$ beams at the ILC can be polarized, with polarization degrees of $|P(e^+)|
= 50$\% and $|P(e^-)| = 80$\%. However, since the realization of positron
polarization poses a serious challenge for the accelerator design, it is
interesting to investigate how much the results presented above depend on it.
As an extreme case, in this section the situation of zero positron polarization
is studied, while as before 80\% polarization is assumed for the electron beam,
$|P(e^+)|$/$|P(e^-)|$ = $0$\%/$80$\%.
The availability of $e^-$ polarization ensures that all observables that are
considered in the previous sections for the ILC still exist in this case. In
particular, supersymmetric production cross-sections can be measured for two
different polarization values, $P(e^-) = \pm 80$\%. However, a loss of accuracy
can result from the absence of positron polarization.

In the baryogenesis scenario BGEN, the uncertainty of the determination of the
electroweak phase transition strength and of the dark matter density is
dominated by the error in the stop mass. The stop mass can be extracted from
measuring the stop production cross-section for two different beam polarization
combinations. Including 50\% positron polarization \cite{stop}, an error of
$\delta \mst = 1.2$ GeV is found with this method,  while without positron
polarization, the error is about 20\% larger, $\delta \mst = 1.4$ GeV.
Similarly, the accuracy for the chargino and neutralino parameters $M_1$, $M_2$
and $\mu$ is reduced by roughly 20\% with respect to the values in
eq.~\eqref{eq:inobgen} when positron polarization is absent, but these
parameters have a smaller impact on cosmological quantities. As a result of the
larger stop mass error, a bigger uncertainty for the prediction of the dark
matter density is obtained, $\Omega_{\rm CDM} h^2 = 0.109^{+0.042}_{-0.032}$,
compared to $^{+0.030}_{-0.027}$ with 50\% $e^+$ polarization.
As mentioned at the end of section~\ref{sec:dm}, the stop mass can also be
determined more precisely from a second cross-section measurement near the stop
pair threshold. This method is less sensitive to the beam polarization, and
with zero $e^+$ polarization the errors are only slightly larger: $\delta \mst
= 0.32$ GeV and $\Omega_{\rm CDM} h^2 = 0.109^{+0.015}_{-0.012}$, instead of
$\delta \mst = 0.30$ GeV and $\Omega_{\rm CDM} h^2 = 0.109^{+0.013}_{-0.010}$.

For the focus-point scenario LCC2, the low-energy phenomenology is governed by
the chargino and neutralino states. With 0\% instead of 50\% position
polarization, the uncertainty of the parameters in eq.~\eqref{eq:inofcp} is
larger by roughly 20\% to 30\%. The upper bound that can be extracted
for the sneutrino mass from the chargino cross-section measurements gets much
weaker, $\msn{{\rm e}} < 13$ TeV, since the derivation of this parameters is
very delicate and requires high precision.
The larger errors in the chargino and neutralino parameters also translate
into a larger error on the predicted dark matter density of 3.1\%, compared to
2.5\% with 50\% position
polarization. Nevertheless, the possibility to compute the dark matter density
with about 3\% precision from collider data is still a very impressive result.


\section{Conclusions}
\label{sec:concl}

In various theoretical supersymmetry frameworks,  the mass of most or all
scalar fermions is constrained to be very large, of the order of at least a few
TeV. Typical examples are electroweak baryogenesis in the MSSM, focus-point
supersymmetry and split supersymmetry. Focusing on two benchmark points, one
electroweak baryogenesis scenario and one focus-point  scenario, it was studied
how such a supersymmetry scenario can be explored at future colliders. It was
found that the LHC can make discoveries of several superpartner particles in
the focus-point case, while in the
baryogenesis scenario a new physics signal is likely to be observable at LHC,
but the identification of the contributing superpartner particles
is very challenging. In both scenarios, the ILC can perform precision
measurements to determine the properties of these particles and set bounds on
other unobserved states.

The analysis was performed including realistic signal and background
computations and simple evaluations of statistical and systematic experimental
errors for mass and cross-section measurements.  The study was based
on the full MSSM, {\it i.e.} without assuming a
specific mechanism or pattern for supersymmetry breaking parameters.

Using these results, the
cosmological implications for electroweak baryogenesis with light stops and
stop-neutralino co-annihilation were investigated. It turns out that the
collider data helps to elucidate the strength of the electroweak phase
transition, while the manifestation of the CP-violating source responsible for
the baryon asymmetry remains unconstrained. Furthermore, the collider measurements
can be used to compute the relic dark matter density. By determining the stop
and lightest neutralino masses, the stop-neutralino co-annihilation process can
be strongly constrained and the dark matter density predicted with a precision
of the same order as current astrophysical results. Refinements in the
determination of the stop mass can improve this result significantly. 

In the focus-point scenario, it was found that very precise measurements of the
accessible neutralino and chargino states can be performed at the ILC, combining
mass measurements from distributions and threshold scans together with
cross-section measurements. This allows to set constraints on the slepton mass
scale and to compute very accurate predictions for the relic density at
the per-cent level. Similar conclusions also apply for split supersymmetry if
the lightest chargino and neutralino particles have masses of a
few 100 GeV.

In both scenarios, the dependence on positron beam polarization was
investigated. Assuming 80\% $e^-$ polarization, and comparing the case of
50\% $e^+$ polarization to zero $e^+$
polarization, it is found that all elements of the analysis can be performed
similarly, but roughly 20--30\% precision in the relevant supersymmetry
parameters and derived cosmological quantities is lost without $e^+$
polarization.

The present work has been performed using tree-level formulae and
cross-sections for the ILC analysis and the computation of the dark matter
annihilation rate. For the expected experimental precision, however, radiative
corrections are important and introduce a dependence on other supersymmetry
parameters, {\it e.g.} sfermion masses and mixing outside of the stop sector.
Furthermore, no CP-violating phase for the gaugino parameter $M_1$ was included
in this analysis, which might have interesting effects in the neutralino
sector. These issues will be studied in future work.

The present study shows that, even in the challenging case of heavy
supersymmetric scalars, precise cross-relations between collider physics
and cosmological processes can be established in order to elucidate some of
the main unresolved questions in our understanding of the universe.


\appendix

\section*{Appendix: MSSM case study scenarios} 
\label{sec:app}

\paragraph{BGEN: Baryogenesis scenario}
\anc\\
The numerical analysis is based on
a typical MSSM parameter point characterized by the following weak scale values:
\begin{equation}
\begin{aligned}
m^2_{\rm\tilde{U}_3} &= -99^2 \gev^2,	& M_1 &= 118.8 \gev, \\
m_{\rm\tilde{Q}_3} &= 4330 \gev,	& M_2 &= 225 \gev, \\
A_t &= -1100 \gev,	& |\mu| &= 225 \gev,\\
	&		& \phi_\mu &= 0.2, \\
m_{\rm\tilde{Q},\tilde{U},\tilde{D},\tilde{L},\tilde{R}_{1,2}} &= 10 \tev,
			& \tan\beta &= 5, \\
	&		& m_{\rm A^0} &= 800 \gev.
\end{aligned}
\label{eq:scen}
\end{equation}
Due to constraints from  large one-loop sfermion-neutralino and
sfermion-chargino effects to the electric dipole moments of the electron and
neutron, the  sleptons and squarks of the first two generations are chosen to
be heavy. The masses of the sbottoms and staus are not specified since they
have no relevance for the baryogenesis scenario.

The chosen parameters are compatible with a strongly first order electroweak
phase transition for electroweak baryogenesis, $v(T_{\rm c})/T_{\rm c} \gesim
1$~\cite{CQW,Carena:1997ki}, generate a sufficiently large baryon asymmetry,
$\eta \sim 0.6 \times 10^{-10}$, and yield a value for the dark matter relic
abundance\footnote{The relic dark matter density has been computed with the
code used in Ref.~\cite{morr}.} of $\Omega_{\rm CDM} h^2 = 0.109$, well
within the WMAP bounds. Note that the stop-neutralino co-annihilation mechanism
is effective for the evolution of the dark matter density in this scenario.
Furthermore, the stop
parameters are chosen such that the mass of the lightest Higgs boson is
$m_{\rm h^0} = 117.3$ GeV,  to satisfy the bound from direct searches at
LEP $m_{\rm h^0} \gesim 114.4 \gev$ \cite{lephbound}. It was checked that the
minimum of the scalar potential is color conserving \cite{Carena:1997ki}. 
At tree-level the following masses are obtained for the
relevant supersymmetric particles:
\begin{equation}
\begin{aligned}
m_{\tilde{t}_1} &= 122.5  \gev, & \mneu{1} &= 106.6 \gev, &
\mneu{3} &= 231.2 \gev, & \mcha{1} &= 162.7 \gev, \\
m_{\tilde{t}_2} &= 4333 \gev,	& \mneu{2} &= 170.8 \gev, &
\mneu{4} &= 297.7 \gev,	& \mcha{2} &= 296.2 \gev, \\
\cos \theta_{\tilde{t}} &= 0.010.
\end{aligned}
\end{equation}

\paragraph{LCC2: Focus-point scenario}
\anc\\
The point LCC2 is chosen as a point with sizeable gaugino-Higgsino mixing,
allowing large neutralino annihilation cross-sections into vector bosons. It
was studied previously in Refs.~\cite{FP,peskin}.
The scenario is defined by mSUGRA parameters at the unification scale,
\begin{equation}
m_0 = 3280 \gev, \quad 
M_{1/2} = 300 \gev, \quad
A_0 = 0, \quad
{\rm sign}(\mu) = +, \quad
\tan\beta = 10.
\end{equation}
Since the evolution of parameters in the focus-point region sensitively depends on
the top quark mass $\mt$, it is fixed to the value $\mt = 175$ GeV.
With {\sc Isajet~7.69} \cite{isajet}, the weak scale soft
breaking parameters are obtained as follows:
\begin{equation}
\begin{aligned}
m_{\rm\tilde{U}_3} &= 1969 \gev,	& M_1 &= 123.1 \gev, \\
m_{\rm\tilde{Q}_3} &= 2710 \gev,	& M_2 &= 237.6 \gev, \\
m_{\rm\tilde{D}_3} &= 3240 \gev,	& |\mu| &= 178.6 \gev,\\
m_{\rm\tilde{L}_3} &= 3268 \gev,	& \tan\beta &= 10, \\
m_{\rm\tilde{R}_3} &= 3252 \gev,	& m_{\rm A^0} &= 3242 \gev,\\
m_{\rm\tilde{Q},\tilde{U},\tilde{D},\tilde{L},\tilde{R}_{1,2}} &\sim 3300 \gev.
\end{aligned}
\label{eq:scen2}
\end{equation}
For the relevant supersymmetric particle masses, {\sc Isajet~7.69} gives
\begin{equation}
\begin{aligned}
\mneu{1} &= 107.7 \gev, &
\mneu{3} &= 190.0 \gev, & \mcha{1} &= 159.4 \gev, & m_{\tilde{g}} &= 850 \gev,
\\
\mneu{2} &= 166.3 \gev, &
\mneu{4} &= 294.4 \gev,	& \mcha{2} &= 286.8 \gev. \\
\end{aligned}
\end{equation}
Following the procedure in Ref.~\cite{peskin},
the relic density is computed with {\sc DarkSUSY~4.1} \cite{darksusy} to
$\Omega_{\rm CDM} h^2 = 0.109$.


\section*{Acknowledgments}

The authors are grateful to C.~Bal\'azs, E.~Baltz, O.~Kittel, S.~Kraml and
T.~Plehn for useful discussions. This work was supported by the
Schweizer Nationalfonds and
by Fermilab, operated by Universities Research Association Inc. under contract
no. DE-AC02-76CH03000 with the DOE. A.F. is grateful  for
hospitality at Fermilab, where part of this work was completed.


\end{document}